\documentclass[
12pt,
preprint,preprintnumbers,nofootinbib,
groupedaddress,superscriptaddress,amsmath,amssymb]{revtex4}
\usepackage{graphicx}
\usepackage{dcolumn}
\usepackage{bm}
\usepackage{amssymb}
\usepackage{amsmath}
\usepackage{epsfig}    
\usepackage{color}
\usepackage{slashed}
\usepackage{hhline}

\def\be{\begin{equation}}
\def\ee{\end{equation}}
\newcommand{\bea}{\begin{eqnarray}}
\newcommand{\eea}{\end{eqnarray}}
\newcommand{\nn}{\nonumber}

\numberwithin{equation}{section}

\begin{document}

\title{Neutrino mass with large $SU(2)_L$ multiplet fields}
\preprint{KIAS-P17064}
\author{Takaaki Nomura}
\email{nomura@kias.re.kr}
\affiliation{School of Physics, KIAS, Seoul 02455, Korea}

\author{Hiroshi Okada}
\email{macokada3hiroshi@cts.nthu.edu.tw}
\affiliation{Physics Division, National Center for Theoretical Sciences, Hsinchu, Taiwan 300}

\date{\today}

\begin{abstract}
 We propose an extension of the standard model introducing large $SU(2)_L$ multiplet fields which are quartet and septet scalars and quintet Majorana fermions. These multiplets can induce the neutrino masses via interactions with the $SU(2)$ doublet leptons. We then find the neutrino masses are suppressed by small vacuum expectation value of the quartet/septet and an inverse of quintet fermion mass relaxing the Yukawa hierarchies among the standard model fermions. We also discuss collider physics at the Large Hadron Collider considering production of charged particles in these multiplets, and due to effects of violating custodial symmetry, some specific signatures can be found. Then we discuss the detectability of these signals.
\end{abstract} 
\maketitle
\newpage

\section{Introduction}

 The standard model (SM) includes fields with only $SU(2)_L$ singlet and doublet. However larger $SU(2)_L$ multiplet fields can also be introduced as exotic field contents which would work to explain a mystery in the SM such as non-zero neutrino masses and give rich phenomenologies.~\footnote{{In refs.~\cite{Cirelli:2005uq, Hambye:2009pw, Sierra:2016qfa} they discuss possibilities of dark matter candidate with multiplet fields with generic way, although this issue is beyond our scope.}} 
A large $SU(2)_L$ multiplet scalar field, which is greater than the doublet fundamental representation, can be distinguished from the SM Higgs and/or multi-doublet Higgs models in recent experiments such as the large hadron collider (LHC), by measuring a specific interaction between a charged scalar bosons in a component of the multiplet and  the  two massive physical vector bosons in the SM.
{For example,  Refs.~\cite{Alvarado:2014jva, Geng:2014oea, Harris:2017ecz, Sierra:2016rcz} have discussed minimal models with a septet boson of $Y=2$ that preserves  $\rho=0$ at tree level, and analyzed Higgs phenomenologies, while Ref.~\cite{Nomura:2016jnl} have also accommodated the neutrino masses at one-loop level.}
This kind of term is induced as a consequence of violating the custodial symmetry~\footnote{There is an exception that a septet scalar field with hypercharge $Y=2$ has such a specific interaction despite of violating the custodial symmetry at tree-level.}.
On the other hand, the custodial symmetry is in very good agreement with the recent electroweak precision test~\cite{Agashe:2014kda}, and a stringent constraint on vacuum expectation value (VEV) of the large multiplet scalar has to be imposed. Its maximum VEV should be of the order 1-10 GeV at most. Thus Yukawa hierarchies between the other SM fermions and neutrinos can be rather mild, once one can explain the neutrino masses via such a small VEV after spontaneous electroweak symmetry breaking. 
Thus it would suggest that such a large multiplet scalar could be in favor of generating the neutrino masses.~\footnote{A typical example is type-II seesaw neutrino mass scenario, in which a $SU(2)_L$ triplet boson is introduced~\cite{Magg:1980ut,Konetschny:1977bn}.}

In this paper, we introduce one quartet and one septet scalar fields and quintet Majorana fermions.
 The interaction among the SM $SU(2)_L$ doublet leptons and these large multiplets can induce neutrino masses which are suppressed by small VEVs of quartet and/or septet as well as an inverse of a quintet fermion mass which is TeV scale.
As a result, the scale of neutrino Yukawa coupling can  be reached at the typical scale of muon Yukawa coupling in the SM.
Then we discuss the collider physics considering the production of charged particles in the multiplets and show the detectability at the LHC.

This paper is organized as follows.
In Sec.~II, we introduce our model, derive some formula of active neutrino mass matrix, and show the typical order of Yukawa couplings and related masses.
In Sec.~III, we discuss implications to physics at the LHC focusing on pair production of  charged particles in the multiplets.
We conclude and discuss in Sec.~IV.

\section{ Model setup}
 \begin{widetext}
\begin{center} 
\begin{table}
\begin{tabular}{|c||c|c|c||c|c|c|c|}\hline\hline  
&\multicolumn{3}{c||}{Lepton Fields} & \multicolumn{3}{c|}{Scalar Fields} \\\hline
& ~$L_L$~ & ~$e_R^{}$~& ~$\Sigma_R$ ~ & ~$H$ ~ & ~$\Phi_7$~  & ~$\Phi_4$ \\\hline 
$SU(2)_L$ & $\bm{2}$  & $\bm{1}$  & $\bm{5}$ & $\bm{2}$ & $\bm{7}$ & $\bm{4}$  \\\hline 
$U(1)_Y$ & $-\frac12$ & $-1$  & $0$  & $\frac12$ & ${1}$ & $\frac12$   \\\hline
\end{tabular}
\caption{Contents of fermion and scalar fields
and their charge assignments under $SU(2)_L\times U(1)_Y$.}
\label{tab:1}
\end{table}
\end{center}
\end{widetext}

In this section, we introduce our model and derive some formulas such as the active neutrino mass matrix. 
The particle contents and their charges are shown in Tab.~\ref{tab:1}.
In fermion sector, we add three iso-spin quintet right-handed exotic fermions $\Sigma_R$ which do not have hypercharge.
In scalar sector, an isospin quartet scalar $\Phi_4$ with hypercharge $Y=1/2$ and an isospin septet scalar $\Phi_7$  with hypercharge $Y=1$ are introduced 
in addition to the SM Higgs doublet $H$.
We assume that neutral components of $H$, $\Phi_4$, and $\Phi_7$ have VEVs, which are respectively symbolized by $v/\sqrt2$ and $v_4/\sqrt{2}$ and $v_7/\sqrt{2}$.
These VEVs are constrained by the $\rho$ parameter, which are given by~\cite{Agashe:2014kda}:
\begin{align}
\rho=\frac{v^2+7 v_4^2+22 v_7^2}{v^2+  v_4^2+ 4 v_7^2},
\end{align}
where the experimental bound be within $\rho=1.0004^{+0.0003}_{-0.0004}$ at $2\sigma$ confidential level.
 Requiring the value of $\rho$ is within $2\sigma$ range, we find the bound for several cases;
\begin{align}
v_7 \lesssim  1.53 \ {\rm GeV} \quad (v_7 \gg v_4), \quad v_4 \lesssim  2.65 \ {\rm GeV} \quad (v_4 \gg v_7), \quad v' \lesssim  1.33 \ {\rm GeV} \quad (v' = v_7 = v_4),
\end{align}
where $v_{\rm SM}=\sqrt{v^2+7 v_4^2 + 22 v_7^2} \simeq 246$ GeV is VEV of the SM-like  Higgs. 
We will show small VEVs of $\Phi_4$ and $\Phi_7$ can be obtained naturally and 
the bound $V_{4,7} \lesssim $ 1 GeV is adopted below.

The relevant Lagrangian and Higgs potential under these symmetries are given by
\begin{align}
-\mathcal{L}_{Y}
&=
(y_{\ell})_{ii} \bar L_{L_i} H e_{R_i} +(y_{\nu})_{ij} [\bar L_{L_i} \tilde\Phi_4 \Sigma_{R_j} ]
 +  (M_{R})_i [\bar \Sigma^c_{R_i} \Sigma_{R_i}] + {\rm h.c.}, \\
\mathcal{V}&= -\mu_H^2 H^\dagger H + M_4^2 \Phi_4^\dagger \Phi_4 + M_7^2 \Phi_7^\dagger \Phi_7  + \mathcal{V}_{\rm nontrivial\ term} +\mathcal{V}_4,\\
 \mathcal{V}_{\rm nontrivial\ term}&=
\sum_{A=1-3} \mu_A[\Phi_4 \Phi_7^\dag \Phi_4]_A
+ \lambda_0 [H^\dag \Phi_4^* H H] +{\rm c.c.},
\label{Eq:lag-flavor}
\end{align}
where $i=1-3$, $j=1-3$, $\tau_i(i=1-3)$ is the Pauli matrix, $\tilde\Phi\equiv i\tau_2\Phi^*$, $\mathcal{V}_4$ is the trivial quartic term, $y_\ell$ and $M_R$ can be diagonal basis without loss of generality,
and the term $y_\ell$ generates the SM
charged-lepton masses $m_\ell\equiv y_\ell v/\sqrt2$ after the spontaneous electroweak symmetry breaking by the VEV of $H$.
Note that inside bracket "$[ \ ]$" the $SU(2)_L$ indices are contracted so that it makes singlet where we omit details here.
We work on the basis where all the coefficients are real and positive for simplicity.
{\it Notice here that the two non-trivial terms with $\mu_A$ and $\lambda_0$ forbid a dangerous massless goldstone boson that couples to the SM gauge fields from arising in the theory, by breaking the accidental global $U(1)$ symmetry among bosons.}
The scalar fields can be parameterized as 
\begin{align}
&H =\left[
\begin{array}{c}
w^+\\
\frac{v+h+iz}{\sqrt2}
\end{array}\right],\quad 
\Phi_4 =\left[\varphi^{++}, \varphi_2^{+}, \varphi^0,\varphi_1^{-}\right]^T,
\\
&\Phi_7= 
\left[\phi^{+4},\phi^{+3},\phi_2^{++}, \phi_2^{+},\phi^{0},\phi_1^{-},\phi_1^{--}\right]^T,
\label{component}
\end{align}
where the upper indices of each component represents the electric charges while the lower indices distinguish components with same electric charge, and $\varphi^0\equiv \frac{v_4+\varphi_R+i\varphi_I}{\sqrt2}$, $\phi^0\equiv \frac{v_7+\phi_R+i\phi_I}{\sqrt2}$.
{\it Here we assume that all the mixings are negligible for simplicity}, although
each of component mixes as follows:
The CP-even mass matrix is written in terms of the basis $[h,\varphi_R^0,\phi_R^0]$. 
The CP-odd mass matrix is written in terms of the basis  $[z,\varphi_I^0,\phi_I^0]$, where 
the lightest state is massless that is absorbed by the longitudinal component of the SM $Z$ boson. 
The singly charged mass matrix is written in terms of the basis $[w^+,\varphi_1^{+},\varphi_2^{+}, \phi_1^{+},\phi_2^{+}]$, where 
the lightest state is massless that is absorbed by the longitudinal component of the SM $W^+$ boson. 
The doubly charged mass matrix is written in terms of the basis $[\varphi^{++}, \phi_1^{++},\phi_2^{++}]$.
 Each of the triply and four charged boson is mass eigenstate.
 
 By taking $M_{4,7}^2 > 0$ and applying the conditions $\partial \mathcal{V}/ \partial v_{4,7}=0$, the VEVs of quartet and septet are roughly given by 
\begin{align}
v_4  \sim \frac{\lambda_0 v^3}{M_4^2}, \quad
v_7  \sim \frac{ \tilde \mu v_4^2}{M_7^2}
\end{align}
where $\tilde \mu$ is a linear combination of $\mu_{A}$.
Thus we find that $v_4 \sim 1$ GeV is naturally obtained with $M_4 \sim 1$ TeV and $\lambda_0 \sim 0.1$, and that $v_7$ tends to be smaller than $v_4$, e.g.
$v_7 \sim 10^{-2}$ GeV with $\tilde \mu \sim 10$ TeV, $M_7 \sim 1$ TeV and $v_4 \sim 1$ GeV.

{\it Fermion quintet}:
We define the quintet Majorana fermions as
\begin{align} 
\Sigma_R&
\equiv
\left[\Sigma_1^{++},\Sigma_1^{+},{\Sigma^0}, \Sigma_{2}^-, \Sigma_{2}^{--} \right]_R^T, 
\label{eq:sigmaR}
\end{align}
where the upper indices for components represent the electric charges while the lower indices distinguish components with the same electric charge.
The quintet is also written as $(\Sigma_R)_{ijkl}$ where the indices take $1$ or $2$ corresponding to $SU(2)_L$ doublet index and 
relation between components in Eq.~(\ref{eq:sigmaR}) is given below Eq.~(\ref{eq:sigmaRapp}) in the Appendix.
The mass terms of the components are given by
\begin{align}
M_R [\bar \Sigma_R^c \Sigma_R] &= M_R (\bar \Sigma_R^c)_{ijkl} ( \Sigma_R)_{i'j'k'l'} \epsilon^{ii'} \epsilon^{jj'} \epsilon^{kk'} \epsilon^{ll'} \nonumber \\
& = \bar \Sigma_{1R}^{++ c} \Sigma_{2R}^{--} + \bar \Sigma_{1R}^{+ c} \Sigma_{2R}^{-} + \bar \Sigma_{2R}^{- c} \Sigma_{1R}^{+} + \bar \Sigma_{2R}^{-- c} \Sigma_{1R}^{+} + 
\bar \Sigma_{R}^{0 c} \Sigma_{R}^{0},  
\end{align}
where $\epsilon^{ii'} (i,i' =1,2)$ is the antisymmetric tensor acting on $SU(2)$ representation space. 
Thus $\Sigma_{1}^{\pm (\pm \pm)}$ and $\Sigma_{2}^{\pm (\pm \pm)}$ are combined to make singly(doubly)- charged Dirac fermions while $\Sigma^0_R$ remain as neutral Majorana fermion.
The masses of each component are given by $M_R$ at the tree level where mixing between the SM leptons will be negligibly small.

\subsection{ Neutrino mass matrix}
Let us firstly decompose the relevant Lagrangian down in order to derive the neutrino mass matrix:
\begin{align}
-{\cal L}& \supset (y_\nu)_{ij}
\left[
\bar\nu_{L_i} \left(\frac{1}{\sqrt2}\Sigma_{R_j}^0 \varphi^{0*} +\frac{\sqrt3}{2} \Sigma_{1R_j}^+ \varphi_1^- +  \frac{1}{2}\Sigma_{2R_j}^+ \varphi_{2}^- +  \Sigma_{1R_j}^{++} \varphi^{--}  \right)\right.
\nn\\
&\left. +
\bar\ell_{L_i} \left(\frac{1}{\sqrt2}\Sigma_{R_j}^0 \varphi_1^-  + \frac12 \Sigma_{1R_j}^+ \varphi^{--} +  \frac{\sqrt3}{2}\Sigma_{2R_j}^- \varphi^{0*} +  \Sigma_{2R_j}^{--} \varphi_{2}^{+}  \right)\right] +{\rm h.c.}
\nn\\& \supset 
 \frac{ {(y_\nu)_{ij}}}{\sqrt2}
\bar\nu_{L_i} \Sigma_{R_j}^0 \varphi^{0*} +{\rm h.c.},
\label{eq:Yukawa}
\end{align}
where the terms in the last line contribute to generate the active neutrino masses. 
\begin{figure}[tb]\begin{center}
\includegraphics[width=70mm]{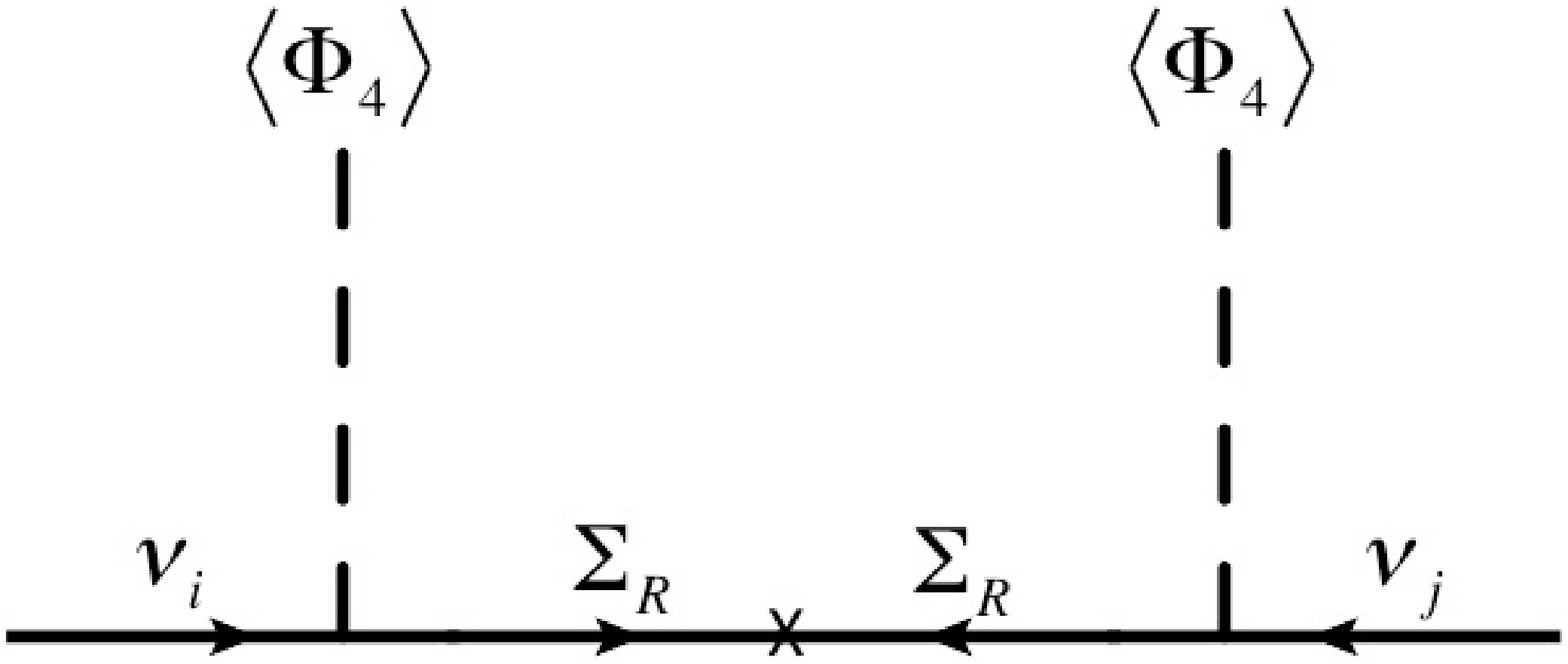}
\includegraphics[width=70mm]{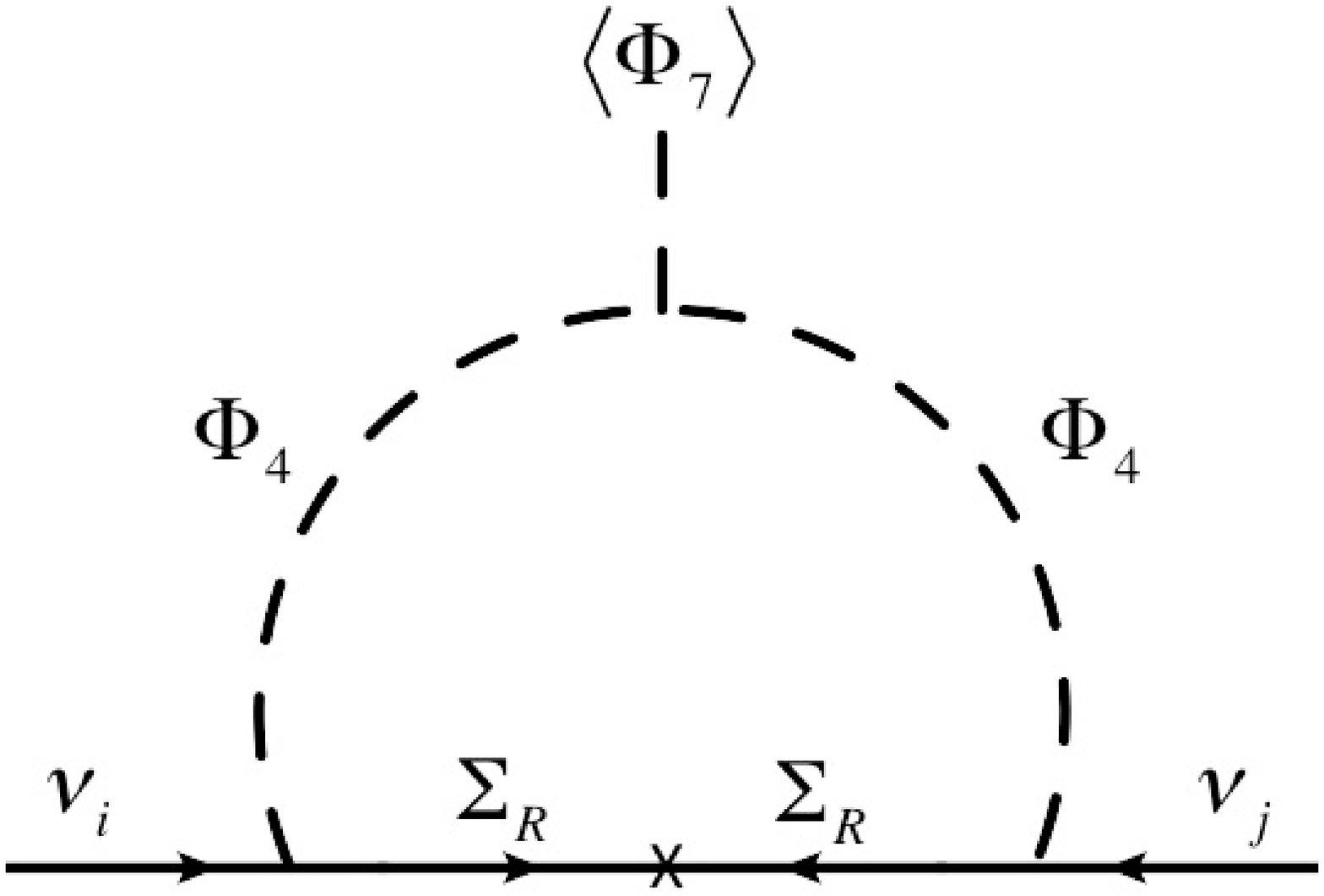}
\caption{Left: Tree level contribution to neutrino masses. Right: Contribution to neutrino masses at one-loop level.
}   \label{fig:neut1}\end{center}\end{figure}
Then the formula of active neutrino mass matrix $m_\nu$ as shown in Figure~\ref{fig:neut1} is given by 
\begin{align}
& (m_{\nu})_{ij}
=\sum_{a=1}^3\frac{(y_\nu)_{ia}(y_\nu^T)_{aj}}{M_{R_a}} \left[v_4^2 + \frac{{\mu v_7}}{(4\pi)^2} F_I(r_{R_a},r_{I_a})\right], \nonumber \\
& F_I(r_1,r_2) =\frac{r_1 \ln[r_1]-r_2\ln[r_2]+r_1r_2\ln[r_2/r_1]}{(1-r_1)(1-r_2)},
\end{align}
where we assume to be $\mu\equiv \mu_1=\mu_2=\mu_3$ for simplicity, and we define $r_{R(I)_a}\equiv (m_{\varphi_{R(I)}}/M_{R_a})^2$ with $a=1-3$.
Note that the first therm in the bracket corresponds to the tree level contribution as the left diagram in Figure~\ref{fig:neut1} while the second term corresponds to the one-loop contribution as the right diagram.
The mass matrix $({m}_\nu)_{ab}$ can be generally diagonalized by the Pontecorvo-Maki-Nakagawa-Sakata mixing matrix $V_{\rm MNS}$ (PMNS)~\cite{Maki:1962mu} as
\begin{align}
({m}_\nu)_{ab} &=(V_{\rm MNS} D_\nu V_{\rm MNS}^T)_{ab},\quad D_\nu\equiv (m_{\nu_1},m_{\nu_2},m_{\nu_3}),
\\
V_{\rm MNS}&=
\left[\begin{array}{ccc} {c_{13}}c_{12} &c_{13}s_{12} & s_{13} e^{-i\delta}\\
 -c_{23}s_{12}-s_{23}s_{13}c_{12}e^{i\delta} & c_{23}c_{12}-s_{23}s_{13}s_{12}e^{i\delta} & s_{23}c_{13}\\
  s_{23}s_{12}-c_{23}s_{13}c_{12}e^{i\delta} & -s_{23}c_{12}-c_{23}s_{13}s_{12}e^{i\delta} & c_{23}c_{13}\\
  \end{array}
\right],
\end{align}
where we neglect the Majorana phase as well as Dirac phase $\delta$ in the numerical analysis for simplicity.
The following neutrino oscillation data at 95\% confidence level~\cite{Agashe:2014kda} is given as
\begin{eqnarray}
&& 0.2911 \leq s_{12}^2 \leq 0.3161, \; 
 0.5262 \leq s_{23}^2 \leq 0.5485, \;
 0.0223 \leq s_{13}^2 \leq 0.0246,  
  \\
&& 
  \ |m_{\nu_3}^2- m_{\nu_2}^2| =(2.44\pm0.06) \times10^{-3} \ {\rm eV}^2,  \; 
  \ m_{\nu_2}^2- m_{\nu_1}^2 =(7.53\pm0.18) \times10^{-5} \ {\rm eV}^2, \nn
  \label{eq:neut-exp}
  \end{eqnarray}
where we assume one of three neutrino masses is zero with normal ordering in our analysis below.
%
The observed PMNS matrix can be realized by introducing the following parametrization.
Here we can parametrize the Yukawa coupling $y_L${, so called Casas-Ibarra parametrization~\cite{Casas:2001sr},} as follows
\begin{align}
y_\nu
&= V_{\rm MNS} \sqrt{D_\nu} OR^{-1/2},
\label{yl-sol}
\\
R_{aa}&\equiv
\sum_{a=1}^3\frac{1}{M_{R_a}}\left[v_4^2 + \frac{{\mu v_7}}{(4\pi)^2} F_I(r_{R_a},r_{I_a})\right],
 \label{R-sol}
\end{align}
where $O$ is an arbitrary complex orthogonal matrix with three degrees of freedom.
Here we estimate the order of Yukawa coupling $y_\nu$. 
First of all, we assume to be $V_{\rm MNS}= O={\cal O}$(1), and $R_{aa}\approx 2 \sum_{a=1}^3\frac{v_4^2}{M_{R_a}}$, where $v_4^2\approx \frac{{\mu v_7}}{(4\pi)^2} F_I(r_{R_a},r_{I_a})$ for $v_4 \sim 1$ GeV, $v_7 \sim 10^{-2}$ GeV and $\mu \sim 10$ TeV taking the values discussed above.
Then the Eq.(\ref{yl-sol}) is simplified as 
\begin{align}
y_\nu= \frac1{\sqrt2 v_4} \sum_{a=1}^3 {\sqrt{D_\nu M_{R_a}}}
\approx
7.2(22.8)\times10^{-4}\ {\rm for}\ \sum_{a=1}^3 M_{R_a}=1(10){\rm TeV},
\label{eq:Yukawa-order}
\end{align}
where we fix to be $D_\nu={\cal O}$(0.1) eV and $v_4$ = 1 GeV.
 Note that the neutrino masses are suppressed by both the small VEVs of $\Phi_4(\Phi_7)$ and TeV scale mass of $\Sigma_R$ giving the order of the Yukawa coupling as $10^{-3}$ which is close to the muon Yukawa coupling in the SM.

{
\subsection{Beta functions of $g$ and $g_Y$}
\label{beta-func}
Here we discuss running of gauge couplings and estimate the effective energy scale by evaluating the Landau poles for $g$ and $g_Y$ in the presence of new fields 
with nonzero multiple hypercharges.
Each of the new beta function of $g$ and  $g_Y$ for one $SU(2)_L$ quintet fermion ($\Sigma_R$), quartet boson ($\Phi_4$),
and septet boson($\Phi_7$) with $(0,1/2,1)$ hypercharge is given by
\begin{align}
\Delta b^{\Sigma_R}_g=\frac{20}{3}, \quad \Delta b^{\Phi_4}_g=\frac{5}{3} \ ,\quad\Delta b^{\Phi_7}_g=\frac{28}{3} ,\\
\Delta b^{\Sigma_R}_Y=0, \quad \Delta b^{\Phi_4}_Y=\frac{3}{5} \ ,\quad\Delta b^{\Phi_7}_Y=\frac{7}{5}.
\end{align}
Then one finds the energy evolution of the gauge coupling $g$ and $g_Y$ as~\cite{Kanemura:2015bli}
\begin{align}
\frac{1}{g^2(\mu)}&=\frac1{g^2(m_{in.})}-\frac{b^{SM}_g}{(4\pi)^2}\ln\left[\frac{\mu^2}{m_{in.}^2}\right]\nn\\
&
-\theta(\mu-m_{th.}) \frac{\Delta b^{\Sigma_R}_g}{(4\pi)^2}\ln\left[\frac{\mu^2}{m_{th.}^2}\right]
-\theta(\mu-m_{th.}) \frac{\Delta b^{\Phi_4}_g+\Delta b^{\Phi_7}_g}{(4\pi)^2}\ln\left[\frac{\mu^2}{m_{th.}^2}\right],\label{eq:rge_g}\\
\frac{1}{g^2_Y(\mu)}&=\frac1{g_Y^2(m_{in.})}-\frac{b^{SM}_Y}{(4\pi)^2}\ln\left[\frac{\mu^2}{m_{in.}^2}\right]
-\theta(\mu-m_{th.}) \frac{\Delta b^{\Phi_4}_Y+\Delta b^{\Phi_7}_Y}{(4\pi)^2}\ln\left[\frac{\mu^2}{m_{th.}^2}\right],\label{eq:rge_gy}
\end{align}
where $\mu$ is a reference energy, $b^{SM}_Y=41/6$, $b^{SM}_g=-19/6$, and we assume to be $m_{in.}(=m_Z)<m_{th.}=$500 GeV, being respectively threshold masses of exotic fermions and bosons for $m_{th.}$.
The resulting flow of $g_Y(\mu)$ is then given by the Fig.~\ref{fig:rge} for each of $g$ and $g_Y$.
This figure shows that $g_Y$ is relevant up to Planck cutoff scale mass, while  $g$ is relevant up to the mass scale $\mu={\cal O}(10^8)$ GeV($=$100 PeV).
Thus our theory does not spoil, as far as we work on at around the scale of 1TeV.

\begin{figure}[tb]
\begin{center}
\includegraphics[width=13cm]{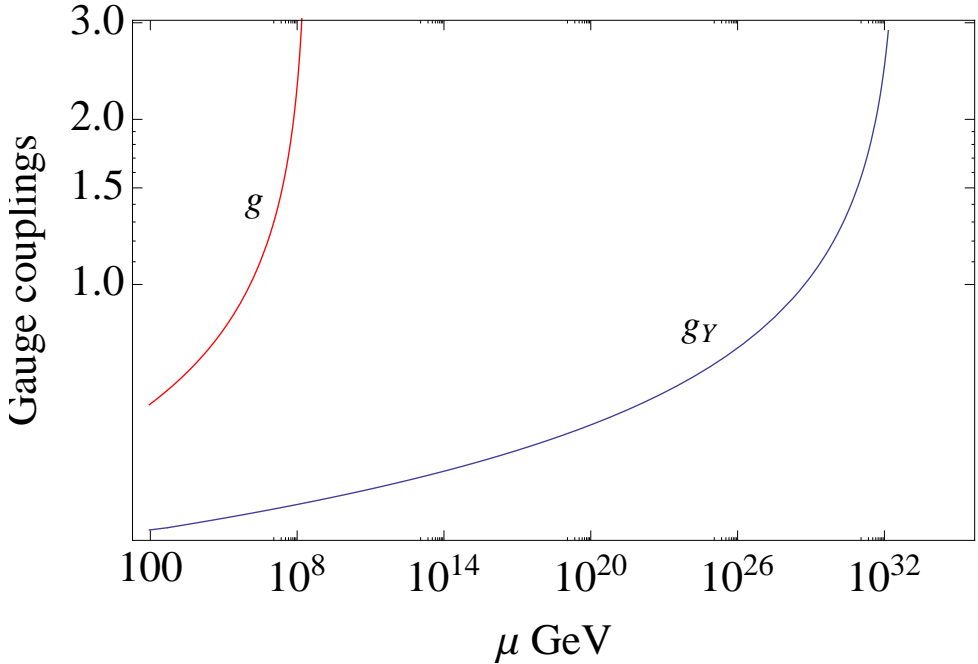}
\caption{The running of $g$ and $g_Y$ in terms of a reference energy of $\mu$,}
\label{fig:rge}
\end{center}\end{figure}

}

\section{Implications to physics at the Large Hadron Collider}

 In this section we discuss signature of our model at the LHC.  We consider pair productions of particles which have the largest electric charge in each multiplets and pair production of singly charged Higgs from the quartet scalar since the former will provide relatively large production cross section and give specific signals while the latter provide $W^\pm Z$ signal indicating the violation of the  custodial symmetry. 
The production cross section for the production processes are estimated using CalcHEP~\cite{Belyaev:2012qa} with {\tt CTEQ6L} PDF~\cite{Nadolsky:2008zw} and applying center of mass energy $\sqrt{s}$ = 13 TeV.  In the following discussion, we consider that components in each multiplet have degenerate mass given by $\{M_4, M_7, M_R \}$ and the case of $M_7 \simeq M_R > M_4$ so that particles in the fermion quintet can decay into scalar bosons in quartet and SM leptons via the Yukawa interaction Eq.~(\ref{eq:Yukawa}).
{Note also that we have running effect for gauge couplings $g$ and $g_Y$ as we discussed above.  The $SU(2)$ coupling $g$ is enhanced by around $10 \%$ to $20 \%$ at $\mathcal{O}(1)$ TeV to $\mathcal{O}(10)$ TeV while the $U(1)_Y$ gauge coupling $g_Y$ is less enhanced. As a result cross sections of exotic particles production will be also enhanced by around $10 \%$ to $40 \%$ level. In our following analysis, we do not consider the running effect for simplicity.}

\subsection{$\varphi^{++} \varphi^{--}$ and $\varphi_1^+ \varphi_1^-(\varphi_2^+ \varphi_2^-)$ pair production as a signature of quartet scalar}

The components of the quartet can be produced via gauge interactions which come from kinetic term after electroweak symmetry breaking;
\begin{align}
|D_\mu \Phi_4|^2 = \sum_{m= -\frac32,-\frac12,\frac12,\frac32} \biggl| & \left[ \partial_\mu - i \left(\frac12+m \right) e A_\mu - i \frac{g}{c_W} \left(m - \left( \frac12+m \right) s_W^2 \right) Z_\mu \right] (\Phi_4)_{m} \nonumber \\
& + \frac{ig}{\sqrt{2}} \sqrt{ \left(\frac32 + m \right) \left(\frac52 -m \right) } W^+_\mu (\Phi_4)_{m-1} \nonumber \\
& + \frac{ig}{\sqrt{2}} \sqrt{ \left(\frac32 - m \right) \left(\frac52 +m \right) } W^-_\mu (\Phi_4)_{m+1} \biggr|^2
\end{align}
where $(\Phi_4)_m$ indicates the component of $\Phi_4$ which has the eigenvalue of diagonal $SU(2)$ generator $T_3$ given by $m$, $g$ is the gauge coupling for $SU(2)_L$, $s_W(c_W) = \sin \theta_W (\cos \theta_W)$ with the Weinberg angle $\theta_W$, and derivation of the covariant derivative for quartet is given in Appendix.
Then we focus on the signatures from $\varphi^{++} \varphi^{--}$ and $\varphi^+_1 \varphi^-_1 (\varphi^+_2 \varphi^-_2)$ production processes which are induced by s-channel $\gamma/Z$ exchange.

In Fig.~\ref{fig:CX}(a), we show the cross sections as functions of quartet mass $M_4$. 
We find that the order of cross section is $\mathcal{O}(1)$ fb for $\mathcal{O}(600)$ GeV scaler mass.
Since we assume the fermion quintets are heavier than the scalar quartet, the produced $\varphi^{\pm \pm}$ and $\varphi^{\pm}_{1,2}$ dominantly decay into $W^\pm W^\pm$ and $W^\pm Z$ respectively through the gauge interaction, 
\begin{equation}
|D_\mu \Phi_4|^2 \supset \sqrt{\frac{3}{2}} v_4 W^\pm W^\pm \varphi^{\mp \mp} + 
\frac{g^2 v_4}{c_W} \left[ s_W^2 Z_\mu W^{+ \mu} \varphi^-_2 + \frac{\sqrt{3}}{2}(2 - s_W^2 )    Z_\mu W^{+ \mu } \varphi^-_1 + c.c. \right].
\end{equation} 
Then we will have the signal processes of 
\begin{align}
& pp \to Z/\gamma \to \varphi^{++} \varphi^{--} \to W^+ W^+ W^- W^-, \\
& pp \to Z/\gamma \to \varphi^{+}_1 \varphi^{-}_1 (\varphi^{+}_2 \varphi^{-}_2) \to W^+ W^- Z Z,
\end{align}
where $W$ and $Z$ bosons further decay into SM fermions. 
For the first process, we consider that two same sign $W$ bosons decay into leptons while the other $W$ bosons decay into jets.
Then the product of cross section and BRs is given by
\begin{equation}
\sigma (pp \to \varphi^{++} \varphi^{--}) BR(W^\pm \to \ell^\pm \nu)^2 BR(W^\pm \to jj)^2 \simeq  0.01 \sigma (pp \to \varphi^{++} \varphi^{--})
\end{equation} 
where $\ell =e, \mu$, $j$ indicate a jet and we have applied $BR(W^\pm \to \ell^\pm \nu) \simeq 0.2$ and $BR(W^\pm \to jj) \simeq 0.7$.
Thus we find that $\mathcal{O}(10)$ events can be produced for $M_4 = 500$ GeV with integrated luminosity 300 fb$^{-1}$ while only a few events are expected for larger mass region. Therefore $\varphi^{\pm \pm}$ with mass of $M_4 \lesssim 500$ GeV would be tested at the LHC by taking appropriate kinematical cuts. 
For the second process, we consider one pair of $W^\pm Z$ decays leptons while the others decay into jets.
Then the product of cross section and BRs is given by
\begin{align}
& [\sigma (pp \to \varphi^{+}_1 \varphi^{-}_1)+\sigma (pp \to \varphi^{+}_2 \varphi^{-}_2)] BR(W^\pm \to \ell^\pm \nu)BR(Z \to \ell^+ \ell^-) BR(W^\pm \to jj) BR(Z \to jj) \nonumber \\
& \simeq  0.007 \times [\sigma (pp \to \varphi^{+}_1 \varphi^{-}_1)+\sigma (pp \to \varphi^{+}_2 \varphi^{-}_2)],
\end{align} 
where we used $BR(Z \to \ell^+ \ell^-) \simeq 0.07$ and $BR(Z \to jj) \simeq 0.7$ in addition to BRs of $W$ boson above.
We then find that $\mathcal{O}(10)$ events can be produced for $M_{4} < 500$ GeV with integrated luminosity 300 fb$^{-1}$.
To estimate discovery significances we need to consider kinematical cuts to reduce SM background. 
However the detailed simulation study of these processes including SM background is beyond the scope of this paper.

\noindent
\subsection{$\phi^{+4} \phi^{-4}$ pair production as a signature of septet scalar}

The components of the septet can be produced via gauge interactions which come from kinetic term after electroweak symmetry breaking;
\begin{align}
|D_\mu \Phi_7|^2 = \sum_{m=-3}^{3} \biggl| & \left[ \partial_\mu - i (1+m) e A_\mu - i \frac{g}{c_W} (m - (1+m) s_W^2) Z_\mu \right] (\Phi_7)_{m} \nonumber \\
& + i\sqrt{ \frac{(3+m)(4-m)}{2} } W^+_\mu (\Phi_7)_{m-1} + i \sqrt{ \frac{(3-m)(4+m)}{2} } W^-_\mu (\Phi_7)_{m+1} \biggr|^2
\end{align}
where $(\Phi_7)_m$ indicates the component of $\Phi_7$ with the eigenvalue of $T_3$ given by $m$ and derivation of the covariant derivative for septet is also given in Appendix.
Then we focus on the signatures from $\phi^{+4} \phi^{-4}$ production process which is induced by s-channel $\gamma/Z$ exchange.

\begin{figure}[tb]\begin{center}
\includegraphics[width=70mm]{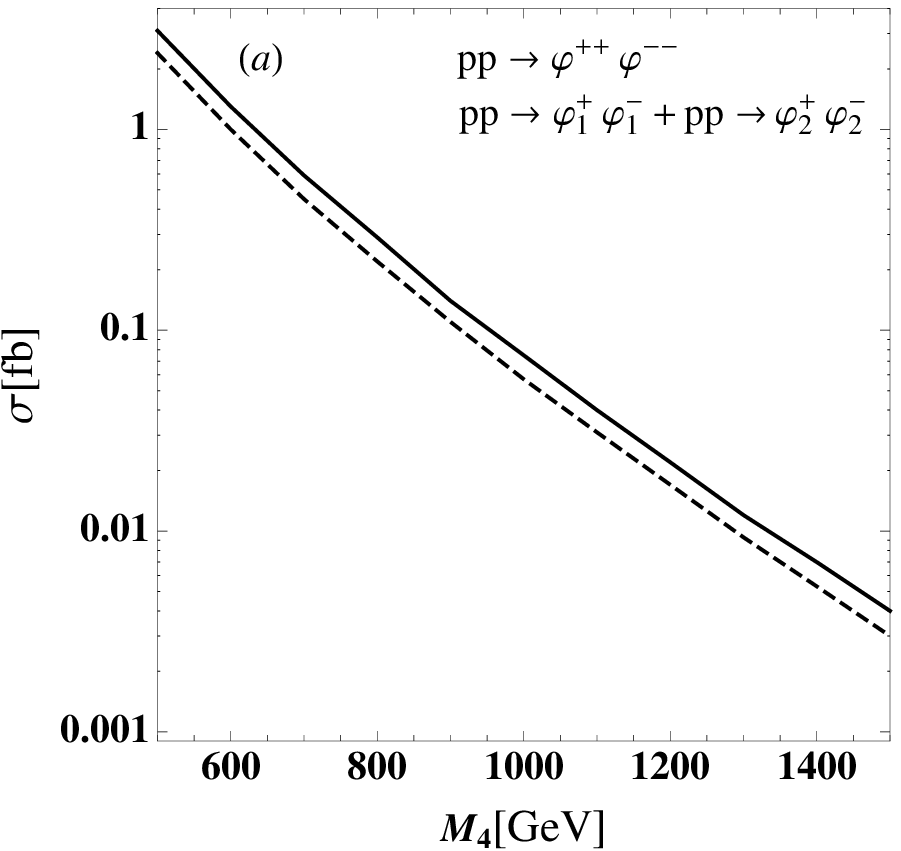}
\includegraphics[width=70mm]{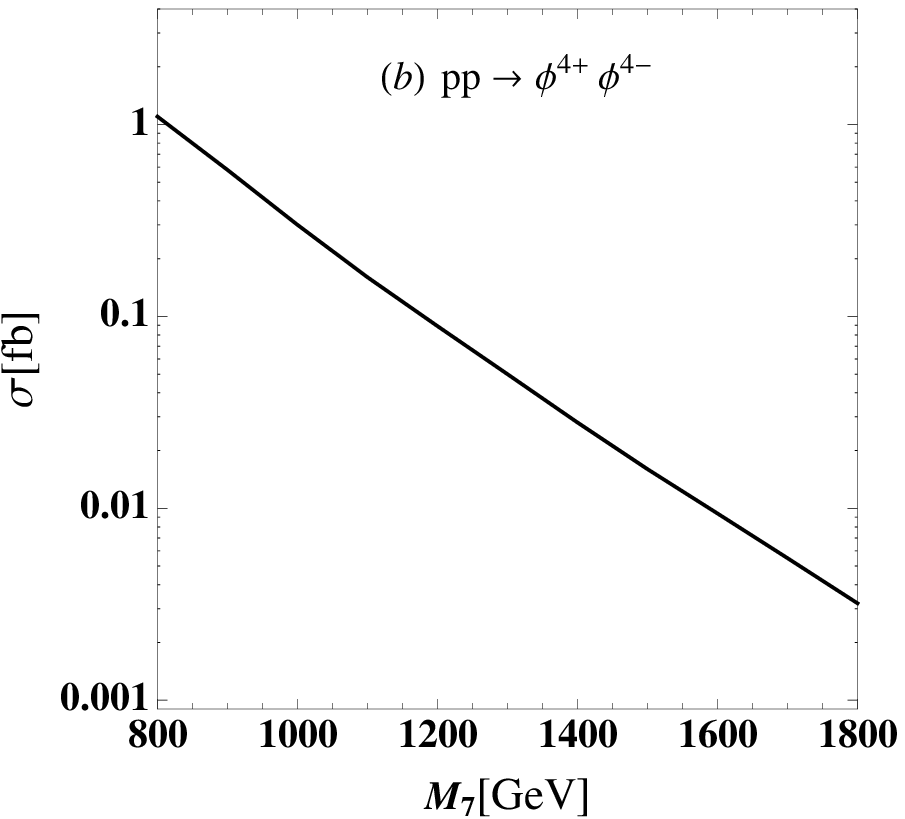}
\caption{(a)The cross section for pair production process $p p \to Z/\gamma \to \varphi^{++} \varphi^{--}$ and $pp \to Z/\gamma \to \varphi^+_1 \varphi^-_1 + pp \to Z/\gamma \to \varphi^+_2 \varphi^-_2$ as a function of quartet mass $M_4$. (b)The cross section for pair production process $p p \to Z/\gamma \to \phi^{+4} \phi^{-4}$ as a function of septet mass $M_7$.
}   \label{fig:CX}\end{center}\end{figure}
In Fig.~\ref{fig:CX}(b), we show the cross section as a function of $M_7$. 
We find that the order of cross section is $\mathcal{O}(1)$ fb for $\mathcal{O}(1)$ TeV scaler mass of the septet scalar bosons.
The produced $\phi^{\pm 4}$ dominantly decays into $\varphi^{\pm \pm} \varphi^{\pm \pm}$ through the trilinear coupling among septet and quartet assuming the mode is kinematically allowed.
Then $\varphi^{\pm \pm}$ dominantly decays into $W^\pm W^\pm$ via gauge interaction.
Thus $pp \to \phi^{+4} \phi^{-4}$ process induces final states with 8 $W$ bosons;
\begin{align}
& p p \to Z/\gamma \to \phi^{+4} \phi^{-4}, \nonumber \\
& \phi^{\pm 4} \to \varphi^{\pm 2} \varphi^{\pm 2} \to W^\pm W^\pm W^\pm W^\pm,
\end{align}
where $W$ bosons further decay into SM fermions.
Here we consider signal events as $n_L$ charged leptons $(e, \mu)$ associated with jets.
Then the expected number of events with $n_L$ leptons is given by
\begin{equation}
N_{n_L} \simeq L \times \sigma(pp \to \phi^{+4} \phi^{-4}) _8 C_x BR(W^\pm \to \ell^\pm \nu)^x BR(W^\pm \to jj)^{8-x}
\end{equation}  
where $L$ denotes the integrated luminosity and we applied $BR(\phi^{\pm4} \to \varphi^{\pm \pm}\varphi^{\pm \pm}) \simeq BR(\varphi^{\pm \pm} \to W^\pm W^\pm) \simeq 1$.
Table~\ref{tab:events} shows the expected number of events with $n_L$ leptons for $L = 300$fb$^{-1}$ and $M_7 = 900$ GeV.
We find that number of events with $n_L \geq 4$ is not sufficiently large while $n_L \leq 3$ cases give number of events larger than $\mathcal{O}(10)$.
The estimation of SM background is required for $n_L \leq 3$ including kinematical cuts to obtain discovery potential which is left in future work.

 \begin{widetext}
\begin{center} 
\begin{table}
\begin{tabular}{|c c c c c c|}\hline
& $n_L \geq 2$ \qquad & $n_L \geq 3$ \qquad & $n_L \geq 4$ \qquad & $n_L \geq 5$ \qquad & $n_L \geq 6$ \\ \hline
\# of events & 42. & 19. & 5.2 & 1.2 & $< 1$ \\ \hline
\end{tabular}
\caption{Number of expected signal events containing $n_L$ charged leptons from $\phi^{+4} \phi^{-4}$ pair production with $L = 300$ fb$^{-1}$ and $M_7 = 900$ GeV. }
\label{tab:events}
\end{table}
\end{center}
\end{widetext}

\subsection{$\Sigma^{++} \Sigma^{--}$ pair production as a signature of quintet fermion}

The components of the quintet can be produced via gauge interactions which come from kinetic term after electroweak symmetry breaking;
\begin{align}
\bar \Sigma_R \gamma^\mu D_\mu \Sigma_R = & \sum_{m=-2}^{2} \biggl[ (\bar \Sigma_R)_m \gamma^\mu \left( \partial_\mu - i m e A_\mu -i g c_W m Z_\mu  \right) (\Sigma_R)_m \nonumber \\
& \quad + \frac{ig}{\sqrt{2}} \sqrt{(2+m)(3-m)} (\bar \Sigma_R)_m \gamma^\mu W_\mu^+  (\Sigma_R)_{m-1} \nonumber \\
& \quad + \frac{ig}{\sqrt{2}} \sqrt{(2-m)(3+m)} (\bar \Sigma_R)_m \gamma^\mu W_\mu^-  (\Sigma_R)_{m+1} \biggr].
\end{align}
where $(\Sigma_R)_m$ indicates the component of $\Sigma_R$ with the eigenvalue of $T_3$ given by $m$.
Here we focus on the pair production of doubly charged fermion $pp \to Z/\gamma \to \Sigma^{++} \Sigma^{--}$ as a signature of the quintet fermion.
In this discussion, we only consider one $\Sigma_R$ assuming the others are heavier.

\begin{figure}[tb]\begin{center}
\includegraphics[width=70mm]{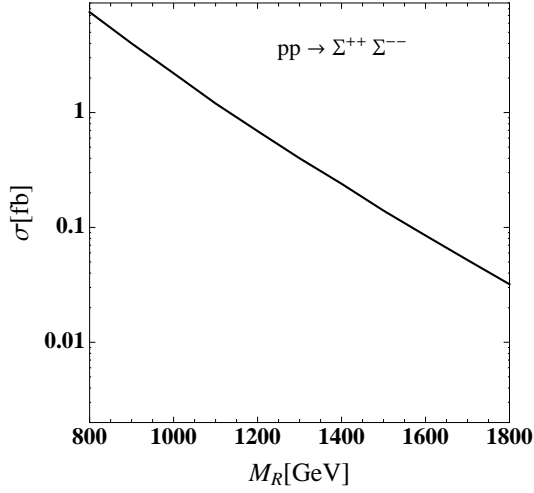}
\caption{The cross section for pair production process $p p \to Z/\gamma \to \Sigma^{++} \Sigma^{--}$ as a function of $\Sigma^{\pm \pm}$ mass.
}   \label{fig:CXsigma}\end{center}\end{figure}
In Fig.~\ref{fig:CXsigma}, we show the cross section as a function of $M_R$. 
The $\Sigma^{\pm \pm}$ then decays as $\Sigma^{\pm \pm} \to \varphi^{\pm \pm} \nu$ and $\Sigma^{\pm \pm} \to \varphi^\pm_2 \ell'^\pm (\ell' = e,\mu, \tau)$ with the same branching fraction through the Yukawa coupling Eq.~(\ref{eq:Yukawa}). Note that $\varphi^{\pm \pm}$ and $\varphi_2^\pm$ further decay into SM gauge bosons as $\varphi^{\pm \pm} \to W^\pm W^\pm$ and $\varphi_2^\pm \to W^\pm Z$.
Thus the the production processes give signals as
\begin{align}
pp \to \Sigma^{++} \Sigma^{--} \to \{W^+ W^+ W^- W^- \nu \nu, W^\pm W^\pm W^\mp \ell^\mp, W^+W^- Z Z \ell^+ \ell^- \}
\end{align}
where these four modes have the same branching ration as $0.25$.
Among these $W^\pm W^\pm W^\mp \ell^\mp$ modes provide the largest number of events since the number of SM gauge bosons is smaller than the others.
Then we consider two same sign $W$ bosons decay into leptons and the other $W$ boson decays into jet.
The product of cross section and BRs is given by
\begin{equation}
\sigma(pp \to \Sigma^{++} \Sigma^{--}) BR(W^\pm \to \ell^\pm \nu)^2 BR(W^\mp \to jj) \sim 0.03 \times \sigma(pp \to \Sigma^{++} \Sigma^{--}).
\end{equation}
When we take $M_R = 800$ GeV and integrated luminosity as 300 fb$^{-1}$ number of produced event will be $\sim 30$ which would be tested by applying kinematical cuts to reduce SM background event. In particular, invariant mass of $\ell^\pm \ell^\pm$ and $\ell^\mp j j$ will be useful to select the signal events which give bump in the distributions according to the mass of the quintet fermion.


\section{ Conclusions and discussions}
 In this paper, we have considered an extension of the SM introducing $SU(2)_L$ quartet and septet scalar fields, and Majorana quintet fermions. 
The quartet and septet scalars have vacuum expectation values which are constrained by the $\rho$-parameter and the smallness can be explained naturally by the parameters in the model.
Then the active neutrino masses can be induced by interactions among these multiplets and the neutrinos.
We have found that the neutrino masses are suppressed by the small VEVs of the quartet/septet and an inverse of TeV scale quintet fermion mass 
explaining the smallness of the neutrino mass with relaxing the Yukawa hierarchies.

We have also discussed the collider physics considering production processes of charged particles in the large multiplets.
The production cross section can be $O(1)$ fb when the mass scale is $\lesssim 1$ TeV.
The charged scalar bosons in the quartet  dominantly decay into $W^\pm$ and $Z$ gauge bosons assuming the quintet fermions are heavier than the quartet scalar.
Then up to $\mathcal{O}(500)$ GeV mass of quartet would be in reach of testing at the LHC experiments.
The scalar with four unit of electric charge in septet decays into doubly charged scalar in quartet and the pair production process gives multiple lepton final states with jets.
We have found that more than $\mathcal{O}(10)$ events can be produced with $900$ GeV mass depending on the number of leptons in final state.
The doubly charged fermion in the quintet also provides multiple lepton final states.
To estimate discovery significance for these signals, the event simulation including the SM background is required which is the beyond the scope of this paper and will be done elsewhere.

{ }


\section*{Acknowledgments}
\vspace{0.5cm}
H. O. is sincerely grateful for all the KIAS members.

\appendix
\section{ Appendix: $SU(2)_L$ large multiplet fields}

\noindent
{\bf Scalar quartet field}

The quartet $\Phi_4$ with hypercharge $Y=1/2$ is represented as 
\begin{equation}
\Phi_4 = \left( \varphi^{++}, \varphi^{+}_2, \varphi^{0}, \varphi^{-}_1 \right)^T, \quad {\rm or} \quad
(\Phi_4)_{ijk}, 
\end{equation}
where $(\Phi_4)_{ijk}$ is the symmetric tensor notation as 
$(\Phi_4)_{[111]} = \varphi^{++}$, $(\Phi_4)_{[112]} = \varphi^{+}_2/\sqrt{3}$, $(\Phi_7)_{[122]} = \varphi^{0}/\sqrt{3}$ and $(\Phi_4)_{[222]} = \varphi^{-}_1$; $[ijk]$ indicates indices are symmetric under exchange among them.
By the expression, we have 
\begin{align}
\Phi_4^\dagger \Phi_4 &=   (\Phi_4^*)_{ijk} (\Phi_4)_{ijk} \nonumber \\
&= \varphi^{++} \varphi^{--} + \varphi^{+}_1 \varphi^{-}_1  + \varphi^{+}_2 \varphi^{-}_2  +  \varphi^0 \varphi^{0} 
\end{align}
where the iterated indices are summed out. 
The covariant derivative of $\Phi_4$ is 
\begin{equation}
D^\mu \Phi_4 = \partial^\mu \Phi_4 - i \left( g W_a^\mu {\cal T}_a^{(4)} +  \frac{1}{2} g' B^\mu \right) \Phi_4,
\end{equation}
where $g(g')$ is the gauge coupling for $SU(2)_L(U(1)_Y)$ and ${\cal T}^{(4)}_a$ is matrices for the generators of SU(2) acting on $\Phi_4^{}$ such that
 \begin{align}
{\cal T}^1 = \frac{1}{2}  \left( \begin{array}{cccc}
   0 & \sqrt{3} & 0 & 0  \\ 
    \sqrt{3} & 0 & 2 & 0  \\ 
    0 & 2 & 0 & \sqrt{3}  \\ 
    0 & 0 & \sqrt{3} & 0 \\ 
  \end{array}  \right)\,, \ \ \ {\cal T}^2 = \frac{i}{2}  \left( \begin{array}{cccc}
   0 & -\sqrt{3} & 0 & 0  \\ 
    \sqrt{3} & 0 & -2 & 0 \\ 
    0 & 2 & 0 & -\sqrt{3} \\ 
    0 & 0 & \sqrt{3} & 0  \\ 
  \end{array}  \right)\,,
 \end{align}
and ${\cal T}^3 = {\rm diag}(3/2, 1/2,  -1/2, -3/2)$.
The covariant derivative in terms of mass eigenstate of SM gauge boson can be obtained by using $W^\pm_\mu = (W_{1 \mu} \mp W_{2 \mu})/\sqrt{2}$, $Z_\mu = \cos \theta_W W_{3 \mu} - \sin \theta_W B_\mu$ and $A_\mu = \sin \theta_W W_{3 \mu} + \cos \theta_W B_\mu$ where $\theta_W$ is the Weinberg angle.  Then we obtain the covariant derivative in terms of mass eigenstates of gauge bosons such that
\begin{align}
(D_\mu \Phi_4)_m =& \left[ \partial_\mu - i \left(\frac12+m \right) e A_\mu - i \frac{g}{c_W} \left(m - \left( \frac12+m \right) s_W^2 \right) Z_\mu \right] (\Phi_4)_{m} \nonumber \\
& + \frac{i}{\sqrt{2}} \sqrt{ \left(\frac32 + m \right) \left(\frac52 -m \right) } W^+_\mu (\Phi_4)_{m-1}  + \frac{i}{\sqrt{2}} \sqrt{ \left(\frac32 - m \right) \left(\frac52 +m \right) } W^-_\mu (\Phi_4)_{m+1} ,
\end{align}
where the subscript $m$ distinguish component of the multiplet by the eigenvalue of ${\cal T}^3$. \\

\noindent
{\bf Scalar quartet field}

The septet $\Phi_7$ with hypercharge $Y=1$ is represented as 
\begin{equation}
\Phi_7 = \left( \phi^{4+}, \phi^{3+}, \phi^{++}_2, \phi^{+}_2, \phi^{0}, \phi^{-}_1, \phi^{--}_1 \right)^T, \quad {\rm or} \quad
(\Phi_7)_{ijklmn}, 
\end{equation}
where $(\Phi_7)_{ijklmn}$ is the symmetric tensor notation as 
$(\Phi_7)_{[111111]} = \phi^{4+}$, $(\Phi_7)_{[111112]} = \phi^{3+}\sqrt{6}$, $(\Phi_7)_{[111122]} = \phi^{++}_2/\sqrt{15}$, $(\Phi_7)_{[111222]} = \phi^{+}_2/\sqrt{20}$, $(\Phi_7)_{[112222]} = \phi^{0}/\sqrt{15}$, $(\Phi_7)_{[122222]} = \phi^{-}_1/\sqrt{6}$ and $(\Phi_7)_{[222222]} = \phi^{--}_1/\sqrt{15}$. 
By the expression, we have 
\begin{align}
\Phi_7^\dagger \Phi_7 =& (\Phi_7^*)_{ijklmn} (\Phi_7)_{ijklmn}  \nonumber \\
= & \phi^{4+} \phi^{4-} + \phi^{3+} \phi^{3-}  + \phi^{++}_1 \phi^{--}_1 + \phi^{++}_2 \phi^{--}_2 + \phi^{+}_1 \phi^{-}_1 + \phi^{+}_2 \phi^{-}_2 +  \phi^0 \phi^{0}, 
\end{align} 
as in the case of quartet.

The covariant derivative of $\Phi_7$ could be expressed by 
\begin{equation}
D^\mu \Phi_7 = \partial^\mu \Phi_7 - i (g W_a^\mu {\cal T}_a^{(7)} +  g' B^\mu ) \Phi_7,
\end{equation}
where ${\cal T}^{(7)}_a$ is matrices for the generators of SU(2) acting on $\Phi_7^{}$ such that
\begin{eqnarray} & \displaystyle
{\cal T}_1^{(7)} \,\,=\,\, \frac{1}{\sqrt{2}}
\begin{pmatrix}
0 & \sqrt{3} & 0 & 0 & 0 & 0 & 0 \\
\sqrt{3} & 0 & \sqrt{5} & 0 & 0 & 0 & 0 \\
0 & \sqrt{5} & 0 & \sqrt{6} & 0 & 0 & 0 \\
0 & 0 & \sqrt{6} & 0 & \sqrt{6} & 0 & 0 \\
0 & 0 & 0 & \sqrt{6} & 0 & \sqrt{5} & 0  \\
0 & 0 & 0 & 0 & \sqrt{5} & 0 & \sqrt{3}  \\
0 & 0 & 0 & 0 & 0 & \sqrt{3} & 0
\end{pmatrix} , \nonumber \\ & \displaystyle
{\cal T}_2^{(7)} \,\,=\,\, \frac{i}{\sqrt{2} }
\begin{pmatrix}
0 & -\sqrt{3} & 0 & 0 & 0 & 0 & 0 \\
\sqrt{3} & 0 & -\sqrt{5} & 0 & 0 & 0 & 0 \\
0 & \sqrt{5} & 0 & -\sqrt{6} & 0 & 0 & 0 \\
0 & 0 & \sqrt{6} & 0 & -\sqrt{6} & 0 & 0 \\
0 & 0 & 0 & \sqrt{6} & 0 & -\sqrt{5} & 0  \\
0 & 0 & 0 & 0 & \sqrt{5} & 0 & -\sqrt{3}  \\
0 & 0 & 0 & 0 & 0 & \sqrt{3} & 0
\end{pmatrix} ,
& \nonumber \\ & \displaystyle
{\cal T}_3^{(7)} \,\,=\,\, {\rm diag}(3,2,1,0,-1,-2,-3) ~.&
\end{eqnarray}
As in the quartet case, the covariant derivative in terms of mass eigenstates of gauge bosons is given by
\begin{align}
(D_\mu \Phi_7)_m = & \left[ \partial_\mu - i (1+m) e A_\mu - i \frac{g}{c_W} (m - (1+m) s_W^2) Z_\mu \right] (\Phi_7)_{m} \nonumber \\
& + i\sqrt{ \frac{(3+m)(4-m)}{2} } W^+_\mu (\Phi_7)_{m-1} + i \sqrt{ \frac{(3-m)(4+m)}{2} } W^-_\mu (\Phi_7)_{m+1} \biggr|^2.
\end{align} 

\noindent
{\bf Fermion quintet field}

The fermion quintet $\Sigma_R$ with hypercharge $Y=0$ is represented as 
\begin{equation}
\Sigma = \left[ \Sigma_1^{++}, \Sigma^{+}_1, \Sigma^{0}, \Sigma^{-}_2, \Sigma_2^{--} \right]_R^T, \quad {\rm or} \quad
(\Sigma_R)_{ ijkl}, 
\label{eq:sigmaRapp}
\end{equation}
where $(\Sigma_R)_{ijkl}$ is the symmetric tensor notation as 
$(\Sigma_R)_{[1111]} = \Sigma_{1R}^{++}$, $(\Sigma_4)_{[1112]} = \Sigma_{1R}^{+}/\sqrt{3}$, $(\Sigma_R)_{[1122]} = \Sigma^{0}_R/\sqrt{3}$, $(\Sigma_R)_{[1222]} = \Sigma^{-}_{2R}$ and $(\Sigma_R)_{[2222]} = \Sigma^{--}_{2R}$.
By the expression, we have 
\begin{align}
\bar \Sigma_R \Sigma_R =&  (\bar \Sigma_R)_{ijkl} (\Sigma_R)_{ijkl}  \nonumber \\
= & \bar \Sigma^{++}_{1R} \Sigma^{++}_{1R} + \bar \Sigma^{+}_{1R} \Sigma^{+}_{1R} + \bar \Sigma^{0}_{R} \Sigma^{0}_{R} + \bar \Sigma^{-}_{2R} \Sigma^{-}_{2R} + \bar \Sigma^{--}_{2R} \Sigma^{--}_{2R}.  
\end{align}
The covariant derivative of $\Sigma_R$ could be expressed by 
\begin{equation}
D^\mu \Sigma_R = \partial^\mu \Sigma_R - i g W_a^\mu {\cal T}_a^{(5)} \Sigma_R,
\end{equation}
where ${\cal T}^{(5)}_a$ is matrices for the generators of SU(2) acting on $\Sigma_R$ given by
\begin{eqnarray} & \displaystyle
{\cal T}_1^{(5)} \,\,=\,\, \frac{1}{2}
\begin{pmatrix}
0 & 2 & 0 & 0 & 0 \\
2 & 0 & \sqrt{6} & 0 & 0 \\
0 & \sqrt{6} & 0 & \sqrt{6} & 0 \\
0 & 0 & \sqrt{6} & 0 & 2 \\
0 & 0 & 0 & 2 & 0
\end{pmatrix} , \hspace{5ex}
{\cal T}_2^{(5)} \,\,=\,\, \frac{i}{2}
\begin{pmatrix}
0 & -2 & 0 & 0 & 0 \\
2 & 0 & -\sqrt{6} & 0 & 0 \\
0 & \sqrt{6} & 0 & -\sqrt{6} & 0 \\
0 & 0 & \sqrt{6} & 0 & -2 \\
0 & 0 & 0 & 2 & 0
\end{pmatrix} ,
& \nonumber \\ & \displaystyle
{\cal T}_3^{(5)} \,\,=\,\, {\rm diag}(2,1,0,-1,-2) ~.
\end{eqnarray}
The covariant derivative in terms of mass eigenstates of gauge bosons is given by
\begin{align}
(D_\mu \Sigma_R)_m = &   \left( \partial_\mu - i m e A_\mu -i g c_W m Z_\mu  \right) (\Sigma_R)_m \nonumber \\
&  + \frac{ig}{\sqrt{2}} \sqrt{(2+m)(3-m)} W_\mu^+  (\Sigma_R)_{m-1} + \frac{ig}{\sqrt{2}} \sqrt{(2-m)(3+m)} W_\mu^-  (\Sigma_R)_{m+1} .
\end{align}

\end{document}